\documentclass[aps]{revtex4}

\usepackage{amsmath,amsfonts}
\usepackage[dvips]{graphicx}
\usepackage{rotating}
 
\begin{document}
 
\title{Which cosmological model with dark energy --- phantom or 
$\Lambda$CDM?}

\author{W{\l}odzimierz God{\l}owski}
\affiliation{Astronomical Observatory, Jagiellonian University,
Orla 171, 30-244 Krak{\'o}w, Poland}
\author{Marek Szyd{\l}owski}
\affiliation{Astronomical Observatory, Jagiellonian University,
Orla 171, 30-244 Krak{\'o}w, Poland}
\affiliation{Complex Systems Research Centre, Jagiellonian University,
Reymonta 4, 30-059 Krak{\'o}w, Poland}


\begin{abstract}
In cosmology many dramatically different scenarios with the past (big bang 
versus bounce) and in the future (de Sitter versus big rip) singularities are 
compatible with the present day observations. This difficulty is called the 
degeneracy problem. We use the Akaike and Bayesian information criteria of 
model selection to overcome this degeneracy and to determine a model with such 
a set of parameters which gives the most preferred fit to the SNIa data. We 
consider seven representative scenarios, namely: the CDM models with the 
cosmological constant, with topological defect, with phantom field, with 
bounce, with bouncing phantom field, with brane and model with the linear 
dynamical equation of state parameter. Applying the model selection 
information criteria we show that AIC indicates the flat phantom model while 
BIC indicates both flat phantom and flat $\Lambda$CDM models. Finally we 
conclude that the number of essential parameters chosen by dark energy models 
which are compared with SNIa data is two. 
\end{abstract}

\maketitle

\section{Introduction}
 
The present observations of the distant supernovae type Ia indicate that the 
Universe is presently accelerating \cite{Riess:1998cb,Perlmutter:1998np} 
due to the presence of some unknown form of energy violating the strong energy 
condition $\rho_{X}+3p_{X}>0$ where $\rho_{X}$ and $p_{X}$ are energy 
density and pressure of dark energy, respectively. While the different 
candidates for dark energy were proposed \cite{Sahni:2002kh,Lahav:2004iy} and 
confronted with observations \cite{Kamenshchik:2001cp,Zhu:2003sq,Sen:2002ss,%
Godlowski:2003pd,Godlowski:2004pt,Puetzfeld:2004df,Biesiada:2004td}, the 
cosmological constant $\Lambda$ and phantom fields 
\cite{Caldwell:1999ew,Carroll:2003st,Hsu:2004vr} violating the weak energy 
condition $\rho_X+p_X>0$ are most popular. Whereas both the cosmological 
constant and phantoms fields, described by the barotropic equation of state 
$p_i=w_i\rho$ ($w_i \le -1$), are negligible in the neighborhood of the 
initial singularity, they dominate late time evolution. In the case of the 
$\Lambda$ domination we obtain the de Sitter state as a global attractor 
in the future and for phantom fields there are a big rip singularity. Both 
the singularities are generic features of their models. In the case of the 
big rip singularity at the some finite time the both scale factor and energy 
density are growing at infinity and the unexpected future singularity of 
density appears \cite{Barrow:1986,Caldwell:2003vq}. Alternatively the universe 
evolved from the Planck region where quantum gravity effects would have 
dominate. In this context quantum cosmology is used as a first approximation 
to the quantum gravity expected effects. Recently the proposal of quantum 
loop cosmology offers possibility of avoiding an initial singularity 
\cite{Bojowald:2001xe,Bojowald:2002nz}. For completeness we also consider the 
brane model \cite{Randall:1999ee,Randall:1999vf,Dabrowski:2002sb} and the 
models with dynamical dark energy satisfying the dynamical equation of state 
$p_X=w_X(z)\rho_X$, where $w_X(z)$ is linearized around $z=0$ (the present 
epoch) \cite{Padmanabhan:2002vv,Choudhury:2003tj}.
 
Let us consider four different prototypes of dynamical behavior which take 
place in the neighborhood of the initial and final singularities. They give 
rise the seven representative scenarios, namely, the CDM model with $\Lambda$ 
($\Lambda$CDM), the CDM model with topological defect (TDCDM), the phantom 
CDM model (PhCDM), the bouncing $\Lambda$CDM model (B$\Lambda$CDM), the 
bouncing phantom CDM model (BPhCDM), the brane $\Lambda$CDM model in Randall 
Sundrum version (Br$\Lambda$CDM), and the model with the dynamical equation 
of state parameter linearized around the present epoch (DEQS). They are 
completed in Table~\ref{tab:1}, where the dependence of the Hubble function 
on redshift $z$ is given together with a number of models parameters. 

\begin{sidewaystable}
\begin{tabular}{c|p{3.7cm}|llc}
\hline
case & name of model    & $\qquad H(z)$& free parameters & $d$ \\ \hline
0  & Einstein-de Sitter & $H=H_0 \sqrt{\Omega_{\text{m},0}(1+z)^3+\Omega_{k,0}(1+z)^2}$ & $H_0,\Omega_{\text{m},0}$ & 2 \\
1  & $\Lambda$CDM       & $H=H_0 \sqrt{\Omega_{\text{m},0}(1+z)^3+\Omega_{k,0}(1+z)^2+\Omega_{\Lambda}}$ & $H_0,\Omega_{\text{m},0},\Omega_{\Lambda}$ & 3 \\
2  & TDCDM              & $H=H_0 \sqrt{\Omega_{\text{m},0}(1+z)^3+\Omega_{k,0}(1+z)^2+\Omega_{T,0}(1+z)}$ & $H_0,\Omega_{\text{m},0},\Omega_{T,0}$ & 3 \\
3a & PhCDM, $w=-\frac{4}{3}$ & $H=H_0 \sqrt{\Omega_{\text{m},0}(1+z)^3+\Omega_{k,0}(1+z)^2+\Omega_{Ph,0}(1+z)^{3\left(1+w\right)}}$ & $H_0,\Omega_{\text{m},0},\Omega_{Ph,0}$ & 3 \\
3b & PhCDM, $w$ -- fitted  &                                                                                               & $H_0,\Omega_{\text{m},0},\Omega_{Ph,0},w$ & 4 \\
4a & B$\Lambda$CDM, $n=6$      & $H=H_0 \sqrt{\Omega_{\text{m},0}(1+z)^3+\Omega_{k,0}(1+z)^2-\Omega_{n,0}(1+z)^n+\Omega_{\Lambda}}$ & $H_0,\Omega_{\text{m},0},\Omega_{n,0},\Omega_{\Lambda}$ & 4 \\
4b & B$\Lambda$CDM, $n$ -- fitted   &                                                                                             & $H_0,\Omega_{\text{m},0},\Omega_{n,0},\Omega_{\Lambda},n$ & 5 \\
5a & BPhCDM, $n=6$             & $H=H_0 \sqrt{\Omega_{\text{m},0}(1+z)^3+\Omega_{k,0}(1+z)^2-\Omega_{n,0}(1+z)^n+\Omega_{Ph,0}(1+z)^{-1}}$ & $H_0,\Omega_{\text{m},0},\Omega_{n,0},\Omega_{Ph,0}$ & 4 \\
5b & BPhCDM, $n$ -- fitted &                                                                                                    & $H_0,\Omega_{\text{m},0},\Omega_{n,0},\Omega_{Ph,0},n$ & 5 \\
6  & Br$\Lambda$CDM, $n=6$     & $H=H_0 \sqrt{\Omega_{\text{m},0}(1+z)^3+\Omega_{k,0}(1+z)^2+\Omega_{BRA,0}(1+z)^n+\Omega_{\Lambda}}$ & $H_0,\Omega_{\text{m},0},\Omega_{BRA,0},\Omega_{\Lambda}$ & 4 \\
7a & $\text{DEQS}$, $w_0=-1$, $p_X=(w_0+w_1z)\rho_X$ &
     $H=H_0 \sqrt{\Omega_{\text{m},0}(1+z)^3+\Omega_{k,0}(1+z)^2+\Omega_{X,0}(1+z)^{3\left(w_0-w_1+1\right)}e^{3w_1z}}$ & $H_0,\Omega_{\text{m},0},\Omega_{X,0},w_1$ & 4 \\
7b & $\text{DEQS}$, $w_0=\text{fitted}$, $p_X=(w_0+w_1z)\rho_X$ &
                                                                                                                          & $H_0,\Omega_{\text{m},0},\Omega_{X,0},w_0,w_1$ & 5 \\
\hline
\end{tabular}
\caption{The Hubble function versus redshift for the seven evolutional 
scenarios of the FRW models with dark energy.}
\label{tab:1}
\end{sidewaystable}

We consider the FRW dynamics in which dark energy is present. The acceleration 
of the Universe is due to the presence of dark energy for which the equation 
of state is $p_X=w_X \rho$ and $w_X=\text{const}<0$ (quintessence) or 
$p_X=w_X(z) \rho$ and $w_X$ is variable with the cosmological time scale 
factor or redshift $z$. Therefore the basic equation determining the evolution 
is
\begin{equation}
\label{eq:1}
H^{2} = \frac{\rho_{\text{eff}}}{3}-\frac{k}{a^2}.
\end{equation}
where $\rho_{\text{eff}}(a)$ is effective energy density of noninteracting
``fluids'', $k = \pm 1,0$ is the curvature index. Equation~(\ref{eq:1}) can 
be presented in terms of density parameters
\begin{equation}
\label{eq:2}
\frac{H^{2}}{H_{0}^{2}} = \Omega_{\text{eff}} (z) + \Omega_{k,0}(1+z)^2
\end{equation}
where $\frac{a}{a_0}=\frac{1}{1+z}$,
$\Omega_{\text{eff}}(z)=\Omega_{\text{m},0}(1+z)^3+\Omega_{X,0}f(z)$ and 
$\Omega_{\text{m},0}$ is the density parameter for the (baryonic and dark) 
matter scaling like $a^{-3}$. For $a=a_{0}$ (the present value of the scale 
factor) we obtain the constraint $\Omega_{\text{eff},0} + \Omega_{k,0}=1$.
 
We assumed that energy density satisfies the conservation condition
\begin{equation}
\label{eq:3}
\dot{\rho}_{i} = -3H(\rho_{i} + p_{i}),
\end{equation}
for each component of the fluid $\rho_{\text{eff}}=\Sigma\rho_i$. Then from 
eq.~(\ref{eq:2}) we obtain the constraint relation $\Sigma_i\Omega_{i,0}
+\Omega_{k,0}=1$. In Table~\ref{tab:1} we denote the numbers of parameters 
by $d$ and also presented the names of the free parameters in the models. 
Please noted that for flat model, $\Omega_{k,0}=0$ the number of the models 
parameters is equal $d-1$.

\section{Distant supernovae as cosmological probes of dark energy}

Distant type Ia supernovae surveys allowed us to find that the present Universe 
is accelerating \cite{Riess:1998cb,Perlmutter:1998np}. Every year new SNIa 
enlarge available data sets with more distant objects and lower systematics 
errors. Riess et al. \cite{Riess:2004nr} compiled the latest samples which 
become the standard data sets of SNIa. One of them, the restricted ``Gold'' 
sample of 157 SNIa, is used in our analysis.
 
For the distant SNIa one can directly observe their the apparent magnitude $m$
and redshift $z$. Because the absolute magnitude $\mathcal{M}$ of the
supernovae is related to its absolute luminosity $L$, then the relation
between the luminosity distance $d_L$ and observed $m$ and absolute magnitude 
$M$ has the following form
\begin{equation}
\label{eq:4}
m - M=5\log_{10}d_{L} +25
\end{equation}
Instead of $d_L$, the dimensionless parameter $D_L$
\begin{equation}
\label{eq:5}
D_{L}=H_{0}d_{L}
\end{equation}
is usually used and then eq.~(\ref{eq:4}) changes to
\begin{equation}
\label{eq:6}
\mu \equiv m - M = 5\log_{10}D_{L} + \mathcal{M}
\end{equation}
where
\begin{equation}
\label{eq:7}
\mathcal{M}=-5\log _{10}H_{0}+25.
\end{equation}

We know the absolute magnitude of SNIa from the light curve. The luminosity
distance of supernova can be obtain as the function of redshift
\begin{equation}
\label{eq:8}
d_L(z) =  (1+z) \frac{c}{H_0} \frac{1}{\sqrt{|\Omega_{k,0}|}}
\mathcal{F} \left( H_0 \sqrt{|\Omega_{k,0}|} \int_0^z \frac{d z'}{H(z')} \right)
\end{equation}
where $\Omega_{k,0} = - \frac{k}{H_0^2}$ and
\begin{align}
\label{eq:9}
\mathcal{F}(x) &=  \sinh(x) \quad &\text{for}& \quad k<0  \nonumber \\
\mathcal{F}(x) &=  x       \quad &\text{for}& \quad k=0  \\
\mathcal{F}(x) &=  \sin(x)  \quad &\text{for}& \quad k>0 \nonumber
\end{align}
 
Finally it is possible to probe dark energy which constitutes the main 
contribution to the matter content. It is assumed that supernovae 
measurements come with uncorrelated Gaussian errors and in this case the 
likelihood function $\mathcal{L}$ can be determined from the chi-square 
statistic $\mathcal{L}\propto \exp(-\chi^{2}/2)$ where
\begin{equation}
\label{eq:10}
\chi^{2}=\sum_{i}\frac{(\mu_{i}^{\text{theor}}-\mu_{i}^{\text{obs}})^{2}}
{\sigma_{i}^{2}},
\end{equation}
while the probability density function of cosmological parameters is 
derived from Bayes' theorem \cite{Riess:1998cb}. Therefore, we can perform the 
estimation of model parameters using the minimization procedure, based on 
the likelihood function. Especially we can also derive a one-dimensional 
probability distribution functions (PDFs) for some model parameters.

\begin{table}
\noindent
\caption{The values of AIC and BIC for distinguished models 
(Table \ref{tab:1}) both for flat and non-flat model.}
\label{tab:2}
\begin{tabular}{c|cccc}
\hline \hline
case & AIC ($\Omega_{k,0}=0$) & AIC ($\Omega_{k,0} \ne 0$) 
& BIC ($\Omega_{k,0}=0$) & BIC ($\Omega_{k,0} \ne 0$) \\
\hline
0   & 325.5  & 194.4 & 328.6  & 200.5  \\
1   & 179.9  & 179.9 & 186.0  & 189.0  \\
2   & 183.2  & 180.1 & 189.4  & 194.4  \\
3a  & 178.0  & 179.3 & 184.1  & 188.5  \\
3b  & 178.5  & 179.7 & 187.7  & 191.9  \\
4a  & 181.9  & 181.6 & 191.1  & 193.8  \\
4b  & 183.9  & 183.6 & 196.2  & 198.8  \\
5a  & 180.0  & 181.3 & 189.2  & 193.5  \\
5b  & 182.0  & 183.3 & 194.4  & 198.5  \\
6   & 180.3  & 181.9 & 189.4  & 194.1  \\
7a  & 179.8  & 181.6 & 188.9  & 193.8  \\
7b  & 180.5  & 182.0 & 192.7  & 197.3  \\
\hline
\end{tabular}
\end{table}
 
\begin{figure*}[ht!]
\includegraphics[width=0.3\textwidth]{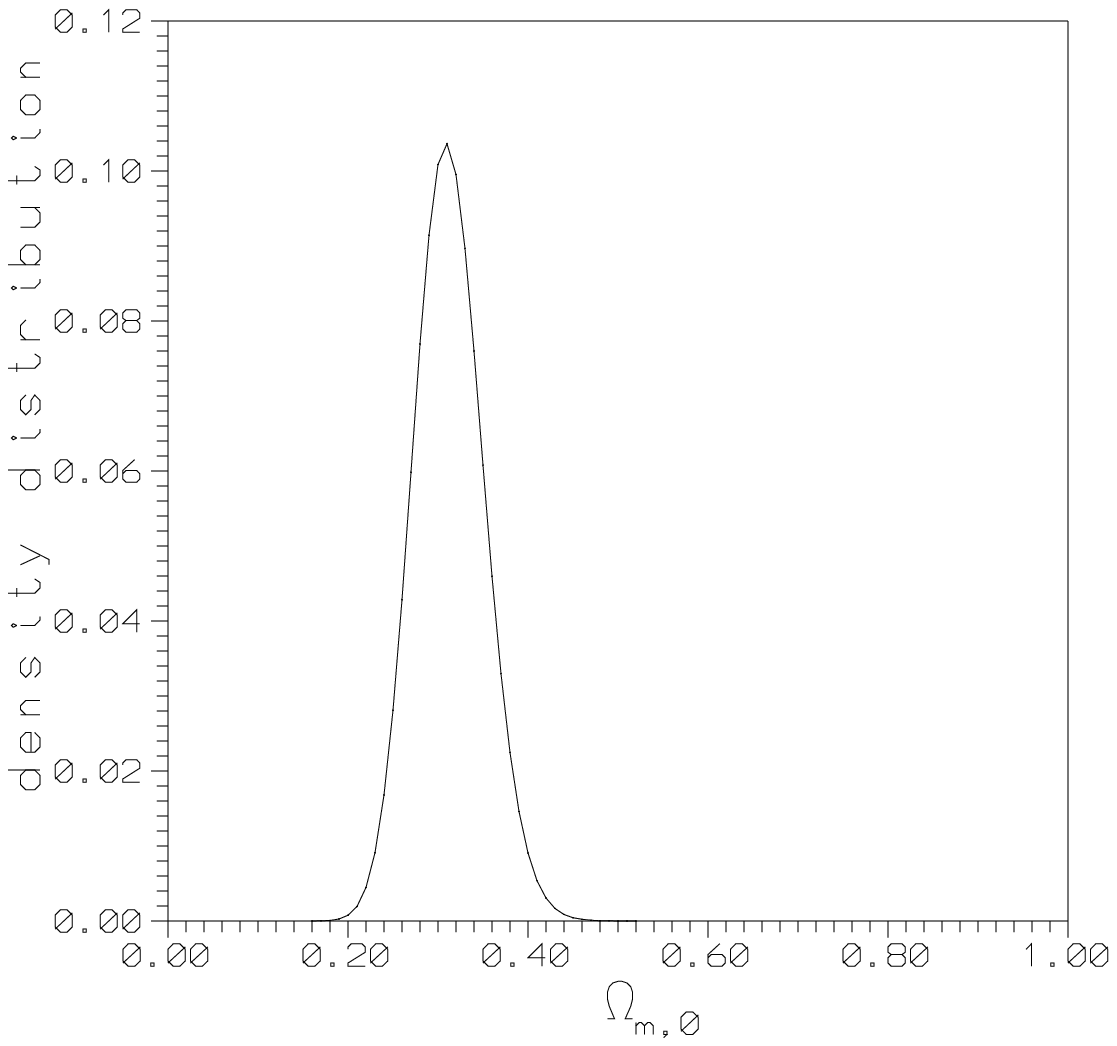}
\includegraphics[width=0.3\textwidth]{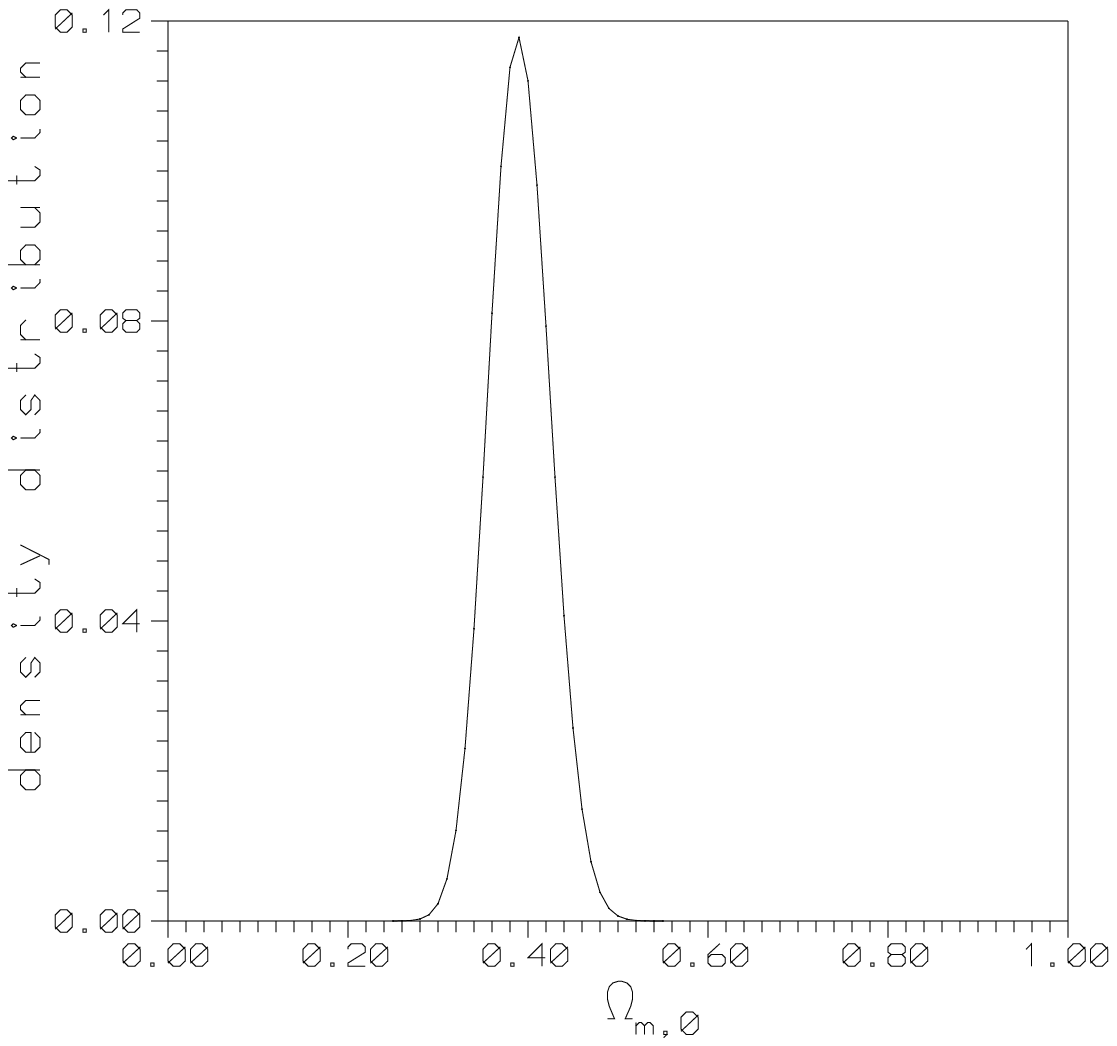}
\includegraphics[width=0.3\textwidth]{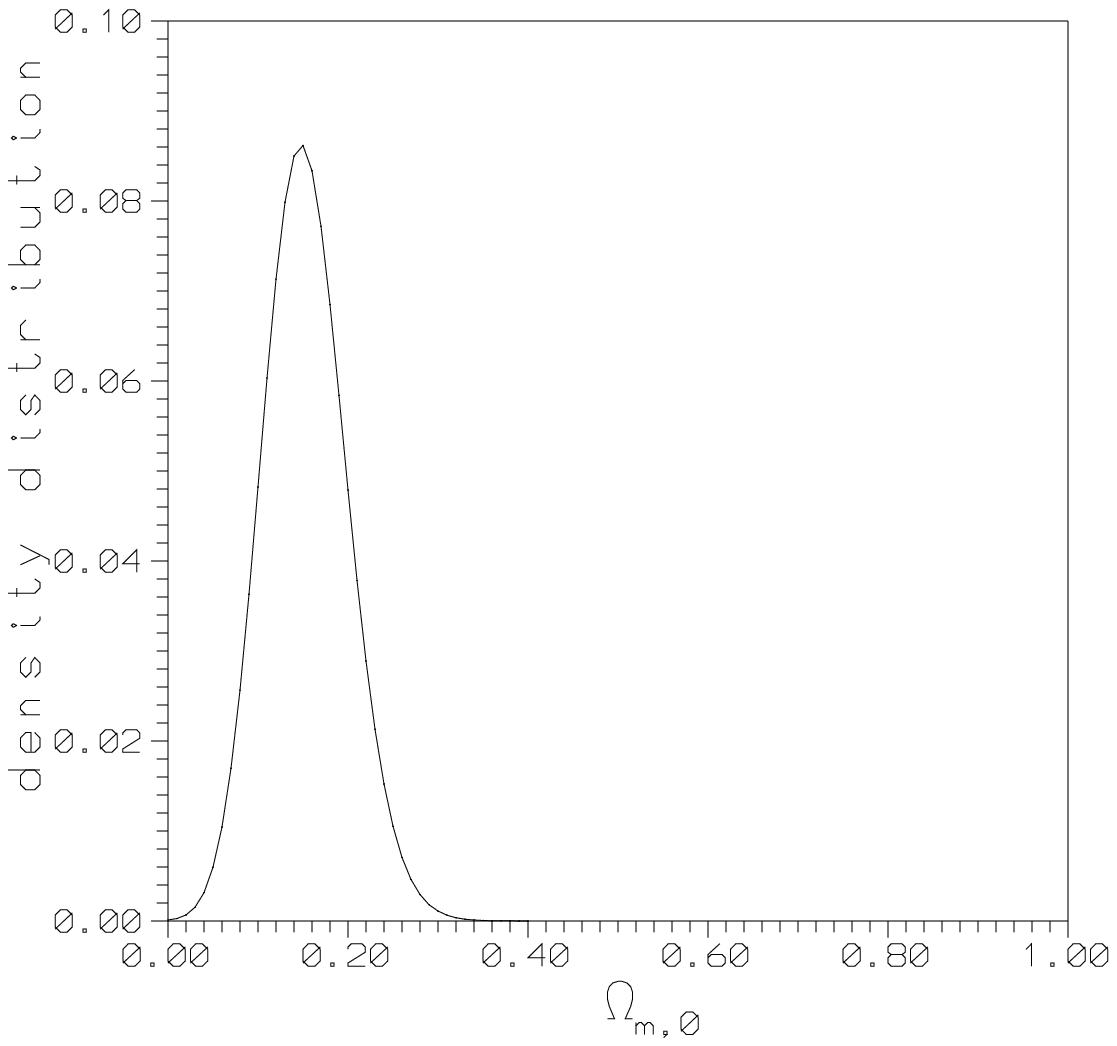}
\caption{The one-dimensional probability density distributions (PDFs) for 
$\Omega_{\text{m},0}$, obtained for (from left to right) the $\Lambda$CDM,
PhCDM, TDCDM models, respectively.}
\label{fig:1}
\end{figure*}

\begin{figure}
\includegraphics[width=0.9\textwidth]{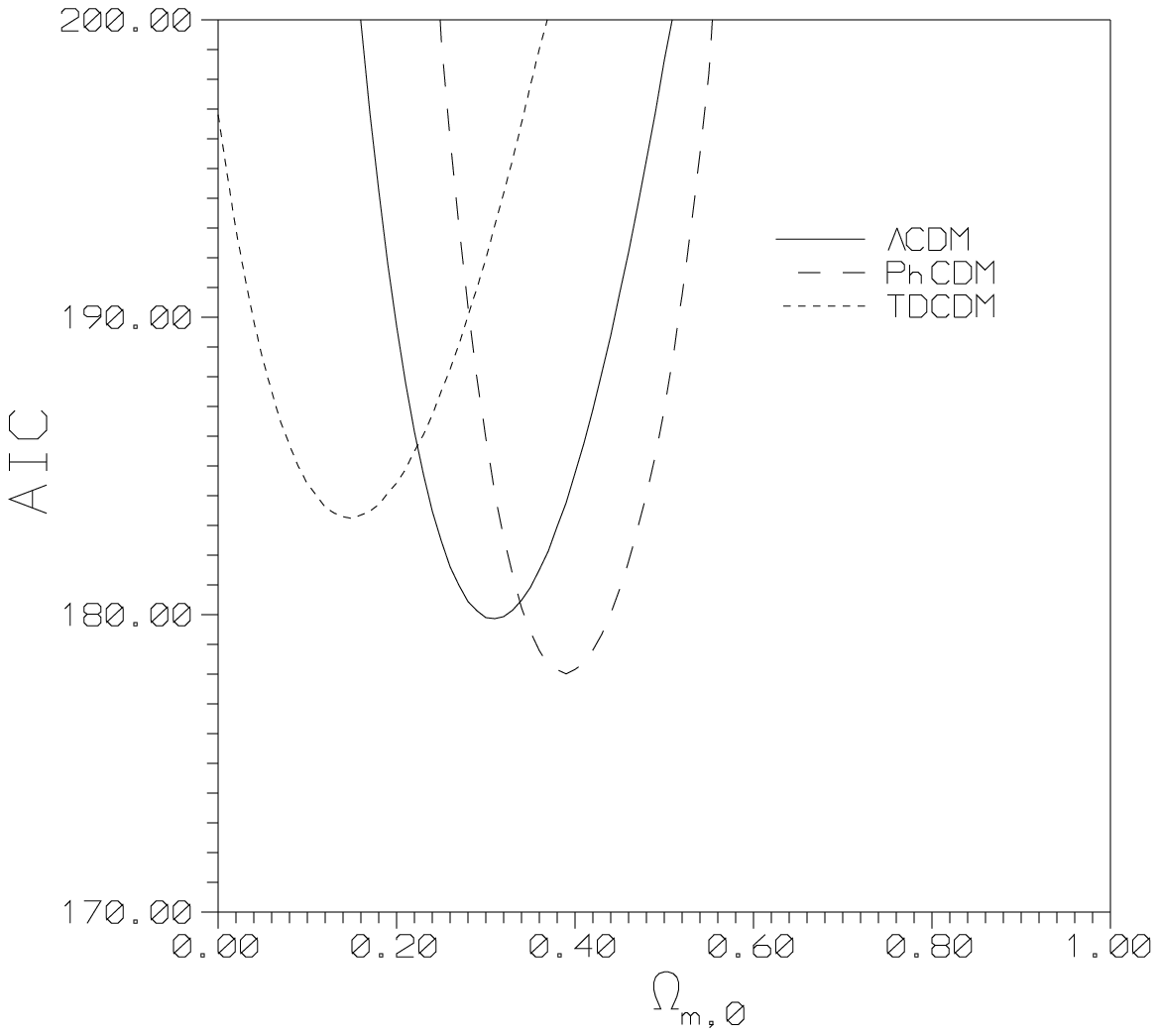}
\caption{The value of AIC in respect to fixed value 
$\Omega_{\text{m},0}$ for three flat models (with  topological defect, with 
cosmological constant and with phantom) with only one parameter $H_0$ 
estimated.}
\label{fig:2}
\end{figure}
 
\begin{figure}
\includegraphics[width=0.9\textwidth]{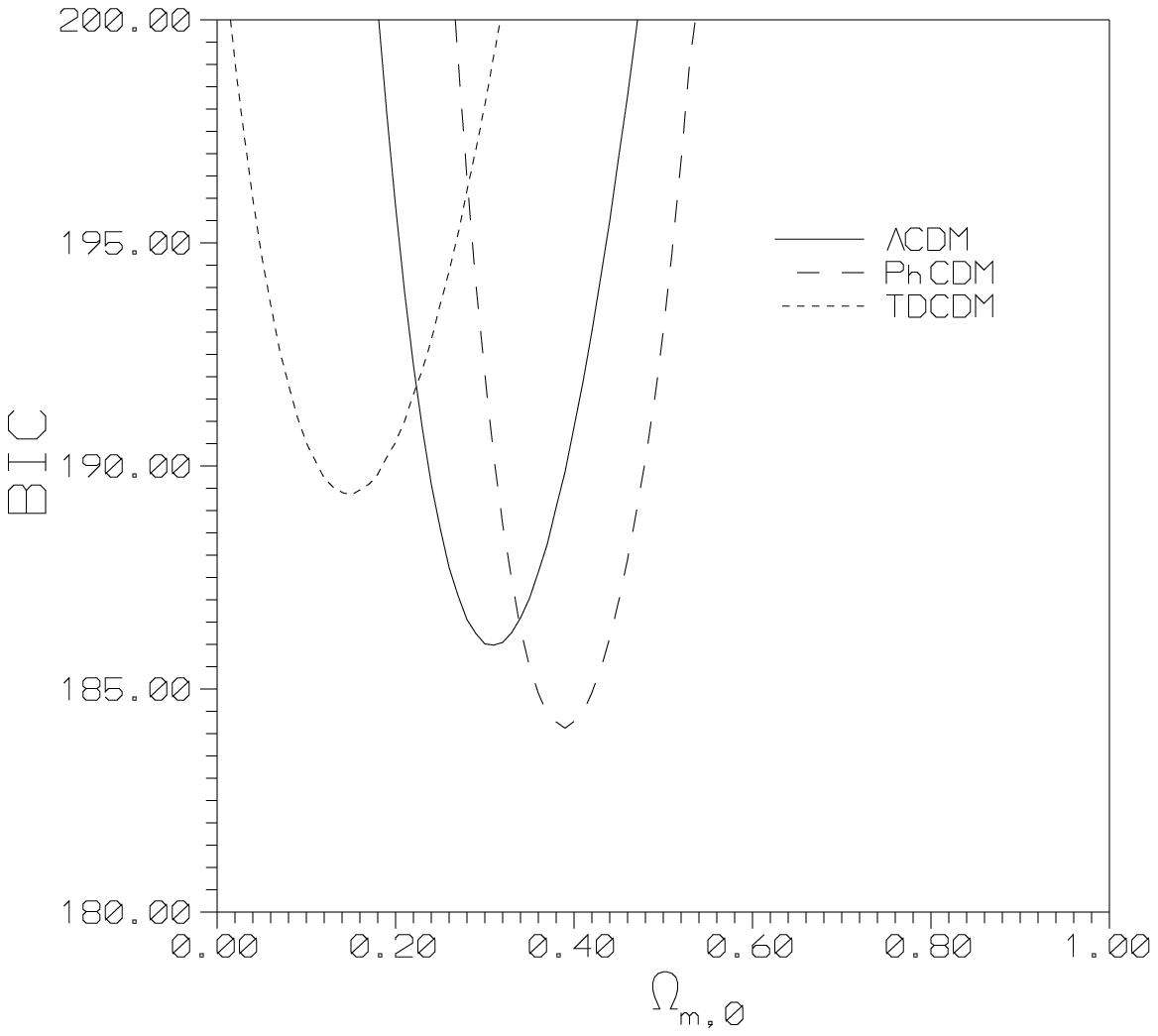}
\caption{The value of BIC in respect to fixed value 
$\Omega_{\text{m},0}$ for three flat models (with topological defect, with 
the cosmological constant and with phantom) with only one parameter $H_0$ 
estimated.}
\label{fig:3}
\end{figure}
\begin{table}
\noindent
\caption{The values of AIC and BIC for distinguished models (Table \ref{tab:1}),
with priors $\Omega_{\text{m},0}=0.3$ both for flat and non-flat model.}
\label{tab:3}
\begin{tabular}{c|cccc}
\hline \hline
case & AIC ($\Omega_{k,0}=0$) & AIC ($\Omega_{k,0} \ne 0$) 
& BIC ($\Omega_{k,0}=0$) & BIC ($\Omega_{k,0} \ne 0$) \\
\hline
0   &  ---   & 216.9 &  ---   & 220.0  \\
1   & 177.9  & 179.9 & 181.0  & 186.0  \\
2   & 190.0  & 178.8 & 193.0  & 184.9  \\
3a  & 183.9  & 179.6 & 187.0  & 186.7  \\
3b  & 179.9  & 178.2 & 186.0  & 187.4  \\
4a  & 179.9  & 181.9 & 186.0  & 191.0  \\
4b  & 181.9  & 183.9 & 191.1  & 196.1  \\
5a  & 185.9  & 181.6 & 192.0  & 190.7  \\
5b  & 187.9  & 183.6 & 197.1  & 195.8  \\
6   & 179.5  & 180.1 & 185.6  & 189.3  \\
7a  & 179.7  & 179.6 & 185.8  & 188.7  \\
7b  & 179.2  & 180.2 & 188.4  & 192.4  \\
\hline
\end{tabular}
\end{table}
 
For deeper statistical analysis we use the Akaike \cite{Akaike:1974} and 
Bayesian information criteria \cite{Schwarz:1978}. It is crucial for our 
investigations to ask how well different theoretical models fits the same 
data set \cite{Liddle:2004nh}. Because from the theoretical investigation 
arising a large number of candidates for dark energy description it is 
necessary to select the essential model parameters from the considered set 
of parameters of the most general model.

The Akaike information criterion (AIC) is defined in the following way 
\cite{Akaike:1974}
\begin{equation}
\label{eq:11}
\text{AIC} = - 2\ln{\mathcal{L}} + 2d
\end{equation}
where $\mathcal{L}$ is the maximum likelihood and $d$ is the number of the 
model parameters. The best model with a parameter set providing the preferred 
fit to the data is that minimizes the AIC.

The Bayesian information criterion (BIC) introduced by Schwarz
\cite{Schwarz:1978} is defined as
\begin{equation}
\label{eq:12}
\text{BIC} = - 2\ln{\mathcal{L}} + d\ln{N}
\end{equation}
where $N$ is the number of data points used in the fit.
 
The effectiveness of using this criteria in the current cosmological 
applications has been recently demonstrated by Liddle \cite{Liddle:2004nh}.
Taking CMB WMAP data \cite{Bennett:2003bz} and applying the information 
criteria, Liddle \cite{Liddle:2004nh} found the number of essential 
cosmological parameters to be five. Moreover he obtained the important 
conclusion that spatially-flat models are statistically preferred to close 
models as it was indicated by the CMB WMAP analysis (their best-fit value is
$\Omega_{\text{tot},0} \equiv \Sigma_i \Omega_{i,0} = 1.02 \pm 0.02$ at 
$1\sigma$ level).

Note that both information criteria have no absolute sense and only the
relative value between differential models are physically interesting.
For the BIC a difference of $2$ is treated as a positive evidence
($6$ as a strong evidence) against the model with larger value of BIC
\cite{Liddle:2004nh,Jeffreys:1961,Mukherjee:1998wp}.
 
The results of calculation AIC and BIC in the context of dark energy models 
are presented in Table~\ref{tab:2} and Table~\ref{tab:3}. In Table~\ref{tab:2} 
we show results for all models considerated for both flat and non flat cases 
without any assumed extra priors. In the general case the number of essential 
parameters in the cosmological models with dark energy is in principal two, 
i.e., ($H_0,\Omega_{\text{m},0}$). It mean that flat model is favored in the 
light of the information criteria. We can observe two rival models which 
minimizes the AIC and BIC. They are the $\Lambda$CDM model and phantom CDM 
(PhCDM) models. One can observe that both BIC and AIC values assume greater 
values for phantom models. This evidence can be regarded as a positive 
evidence in favor of the PhCDM model.

We can derived one-dimensional probability distribution functions (PDFs)
for the models parameters. In Fig.~\ref{fig:1} we present PDFs for 
$\Omega_{\text{m},0}$ for all three flat models with two free parameters, i.e., for the 
flat $\Lambda$CDM, TDCDM and PhCDM models. There is a significant difference 
between predictions of these models. The $\Lambda$CDM model prefers a universe 
with $\Omega_{\text{m},0}$ close to $0.3$, the PhCDM model favors a high 
density universe while the TDCDM model favors a low density universe.
In Fig.~\ref{fig:2} and Fig.~\ref{fig:3} we present values of the AIC and BIC 
for these models. If $\Omega_{\text{m},0}<0.22$ then the information criteria 
favor the TDCDM model. For $\Omega_{\text{m},0} \in (0.22, 0.34)$, the 
$\Lambda$CDM is favored while for $\Omega_{\text{m},0}>0.34$ the PhCDM model 
is preferred.
 
The similar analysis with the use of the information criteria is done in the 
case of the assumed prior $\Omega_{\text{m},0}=0.3$ \cite{Peebles:2002gy}
(Table~\ref{tab:3}).
In this case models with $\Lambda$ is preferred over the models with phantoms,
that is in contrary to the results obtained in the previous case of no priors 
for $\Omega_{\text{m},0}$. It clearly shows that more precise measurements of 
$\Omega_{\text{m},0}$ will give us the possibility to discriminate between 
the $\Lambda$CDM and PhCDM models.

\section{Conclusion}

The main goal of this letter is to decide which model with dark energy is 
distinguish by statistical analysis of SNIa data. To do this we use the 
Akaike and Bayesian information criteria. The former criterion weighs in 
favor of the flat phantom model while the latter distinguishes the flat 
phantom and $\Lambda$CDM models. Assuming the prior $\Omega_{\text{m},0}=0.3$ 
both AIC and BIC criteria weighs in favor of the flat $\Lambda$CDM model. 

It was considered the seven different models containing dark energy and dust 
matter (baryonic and dark matter). Our main result is that spatially flat 
phantom model is favored to $\Lambda$CDM model by information criteria of 
AIC and BIC. The further conclusions are the following.
\begin{itemize}
\item The minimal number of essential parameters in the cosmological models 
with dark energy is in principal two, i.e., ($H_0,\Omega_{\text{m},0}$).
The list of essential parameters may be longer, because some of them are not 
convincingly measured with present data, like the parameter $w_1$.
\item The curvature density parameter does not belong to the class of
essential parameters when all the rest parameters are without any priors 
(with no fixed $\Omega_{\text{m},0}$). At this point our result coincides 
with analogous result obtained by Liddle who found it basing on other 
observations.
\item If we consider models in which all model parameters are fitted then the 
PhCDM model with double initial and final singularities is distinguished.
\item When we consider the prior on $\Omega_{\text{m},0}$, then the model with 
two-dimensional topological defect has $\Omega_{\text{m},0} < 0.22$, 
the $\Lambda$CDM has $\Omega_{\text{m},0} \in (0.22;0.34)$, and the phantom 
model  $\Omega_{\text{m},0} > 0.34$. 
\end{itemize}

To make the ultimate decision which model describes our Universe it is 
necessary to obtain the precise value of $\Omega_{\text{m},0}$ from 
independent observations. 

One can also draw from our analysis some kind of philosophical conclusion. 
Basing on simple and objective information criterion finally we obtain that 
SNIa data favored the models with the initial (big bang) and final (big rip) 
singularities. The big rip singularities are represents generic final future 
state of universe dominating by phantom dark energy. The bounce can be caused
by high energy quantity modifications to FRW equations which make the cosmology 
nonsingular \cite{Szydlowski:2005qb}. If the energy density is so large then 
quantum gravity corrections are important at both the big bang and big rip. 
It is interesting that classical theory reveals itself boundness, i.e., 
classical singularities which can be distinguished by the information criteria.
The account of quantum effects allows to avoid not only the initial
singularity \cite{Bojowald:2001xe,Bojowald:2002nz} but also escape from
the future singularity \cite{Nojiri:2003jn,Elizalde:2004mq,Nojiri:2004pf}.

\acknowledgements{M. Szydlowski acknowledges the support by KBN grant 
no. 1 P03D 003 26.}
 

\end{document}